\documentclass[12pt]{article}

\usepackage{scicite}

\usepackage{times}

\usepackage{graphicx}
\usepackage{wasysym}

\newenvironment{sciabstract}{%
\begin{quote} \bf}
{\end{quote}}

\newcounter{lastnote}
\newenvironment{scilastnote}{%
\setcounter{lastnote}{\value{enumiv}}%
\addtocounter{lastnote}{+1}%
\begin{list}%
{\arabic{lastnote}.}
{\setlength{\leftmargin}{.22in}}
{\setlength{\labelsep}{.5em}}}
{\end{list}}

\title{From a single-band metal to a high-temperature \\superconductor via two thermal phase transitions}

\author
{Rui-Hua He,$^{1, 2, 3\ast}$ M. Hashimoto,$^{1, 2, 3\ast}$ H. Karapetyan,$^{1, 2}$ J. D. Koralek,$^{3, 4}$ \\J. P. Hinton,$^{3, 4}$ J. P. Testaud,$^{1, 2, 3}$ V. Nathan,$^{1, 2}$ Y. Yoshida,$^{5}$ Hong Yao,$^{1, 3, 4}$ \\K. Tanaka,$^{1, 2, 3, 6}$ W. Meevasana,$^{1, 2, 7}$ R. G. Moore,$^{1, 2}$ D. H. Lu,$^{1, 2}$ S.-K. Mo,$^{3}$\\M. Ishikado,$^{8}$ H. Eisaki,$^{5}$ Z. Hussain,$^{3}$ T. P. Devereaux,$^{1, 2\dag}$  \\S. A. Kivelson,$^{1\dag}$ J. Orenstein,$^{3, 4\dag}$ A. Kapitulnik,$^{1, 2\dag}$ Z.-X. Shen$^{1, 2\dag}$\\
\\
\normalsize{$^{1}$Geballe Laboratory for Advanced Materials, Departments of Physics}\\
\normalsize{and Applied Physics, Stanford University, Stanford, California 94305, USA}\\
\normalsize{$^{2}$Stanford Institute for Materials and Energy Sciences,}\\
\normalsize{SLAC National Accelerator Laboratory, Menlo Park, California 94025, USA}\\
\normalsize{$^{3}$Advanced Light Source and Materials Sciences Division,}\\
\normalsize{Lawrence Berkeley National Laboratory, Berkeley, California 94720, USA}\\
\normalsize{$^{4}$Department of Physics, University of California, Berkeley, CA 94720}\\
\normalsize{$^{5}$Nanoelectronics Research Institute, AIST, Ibaraki 305-8568, Japan}\\
\normalsize{$^{6}$Department of Physics, Osaka University, Toyonaka, Osaka 560-0043, Japan}\\
\normalsize{$^{7}$School of Physics, Suranaree University of Technology and}\\
\normalsize{Synchrotron Light Research Institute, Nakhon Ratchasima, 30000 Thailand}\\
\normalsize{$^{8}$Japan Atomic Energy Agency, Tokai, Ibaraki 319-1195, Japan}
\\
\normalsize{$^\ast$These authors contributed equally to this work.}\\
\normalsize{$^\dag$To whom correspondence should be addressed; E-mail: zxshen@stanford.edu,}\\ \normalsize{aharonk@stanford.edu, jworenstein@lbl.gov, kivelson@stanford.edu, tpd@stanford.edu.}
}

\date{}

\begin{document} 


\baselineskip24pt


\maketitle 


\begin{sciabstract}
The nature of the pseudogap phase of cuprate high-temperature superconductors is a major unsolved problem in condensed matter physics. We studied the commencement of the pseudogap state at temperature $T^*$ using three different techniques (angle-resolved photoemission spectroscopy, polar Kerr effect, and time-resolved reflectivity) on the same optimally-doped Bi2201 crystals. We observed the coincident, abrupt onset at $T^*$ of a particle-hole asymmetric antinodal gap in the electronic spectrum, a Kerr rotation in the reflected light polarization, and a change in the ultrafast relaxational dynamics, consistent with a phase transition. Upon further cooling, spectroscopic signatures of superconductivity begin to grow close to the superconducting transition temperature ($T_c$), entangled in an energy-momentum dependent fashion with the pre-existing pseudogap features, ushering in a ground state with coexisting orders.
\end{sciabstract}



As complex oxides, cuprate superconductors belong to a class of materials which exhibit many broken-symmetry states; unravelling the relationship between superconductivity in the cuprates and other possible broken-symmetry states has been a major challenge of condensed matter physics. A possibly related issue concerns the nature of the pseudogap in the cuprates and its relationship with superconductivity. Angle-resolved photoemission spectroscopy (ARPES) studies have shown that the pseudogap develops below a temperature $T^*$ near the Brillouin zone boundary while preserving a gapless Fermi arc near the zone diagonal \cite{misc:PES:ZXreview}. A key issue is the extent to which the pseudogap is a consequence of superconducting fluctuations \cite{HTSC:PrecursorPairing, HTSC:NernstEffect_Ong, cuprates:Bi2201:TakahashiPGSC, cuprates:Bi2201:OneGap_Xingjiang},  which should exhibit a rough particle-hole symmetry, or another form of (incipient) order \cite{stripe:theory:KivelsonStripe_eLiquidXtal, HTSC:CompetingOrders_DDW, stripe:theory:GrilliFluctuatingDW, HTSC:CompetingOrders_VarmaLoops, cuprates:Bi2212:Kiyo_TwoGap, HTSC:Kondo_TwoPGs, cuprates:Bi2201:DingHongTwoGaps}, which typically should induce particle-hole asymmetric spectral changes. Candidate orders include various forms of density wave, nematic or unconventional magnetic orders that break different combinations of lattice translational \cite{stripe:theory:KivelsonStripe_eLiquidXtal, HTSC:CompetingOrders_DDW, stripe:theory:GrilliFluctuatingDW, stripe:neutron:TranquadaStripe1st, stripe:STM:CB_VortexCore, stripe:STM:CB_CB2212_Kapitulnik, stripe:STM:CB_Pseudogap, stripe:STM:C4C2, stripe:STM:PseudogapFluctStripe_Bi2212, stripe:STM:CB_Bi2201}, rotational \cite{stripe:theory:KivelsonStripe_eLiquidXtal, HTSC:CompetingOrders_VarmaLoops, stripe:STM:CB_CB2212_Kapitulnik, stripe:STM:C4C2, stripe:neutron:YBCO_eLiquidCrystal, HTSC:PseudogapBreakRotational, stripe:STM:Nematic_Bi2212}, and time-reversal \cite{HTSC:CompetingOrders_DDW, HTSC:CompetingOrders_VarmaLoops, HTSC:BreakTimeRev_ARPES_Bi2212, HTSC:BreakTimeRev_neutron_YBCO, HTSC:KerrEffect_YBCO, HTSC:BreakTimeRev_neutron_Hg1201} symmetries.

We have focused on crystals of nearly optimally-doped (OP) Pb$_{0.55}$Bi$_{1.5}$Sr$_{1.6}$La$_{0.4}$CuO$_{6+\delta}$ (Pb-Bi2201, $T_c=38$ K, $T^*= 132 \pm 8$ K) \cite{Note_SOM}, and combined the ARPES measurements of the evolution of the band structure over a wide range of temperature, momentum and energy, with high-precision measurements of the polar Kerr effect (PKE) and time-resolved reflectivity (TRR). Bi2201 was chosen to avoid the complications resulting from bilayer splitting and strong antinodal bosonic mode coupling inherent to Bi$_2$Sr$_2$CaCu$_2$O$_{8+\delta}$ (Bi2212) \cite{misc:PES:ZXreview}. Whereas ARPES is a surface probe, PKE enables us to monitor a bulk, thermodynamic (via the fluctuation-dissipation theorem) property which has proven \cite{otheroxides:KerrEffect_Sr2RuO4} to be a sensitive probe of the onset of a broken-symmetry state, and TRR gives complementary information on the bulk, near-equilibrium dynamics of the system.

\begin{figure}
\centering
\includegraphics [width=4.55in]{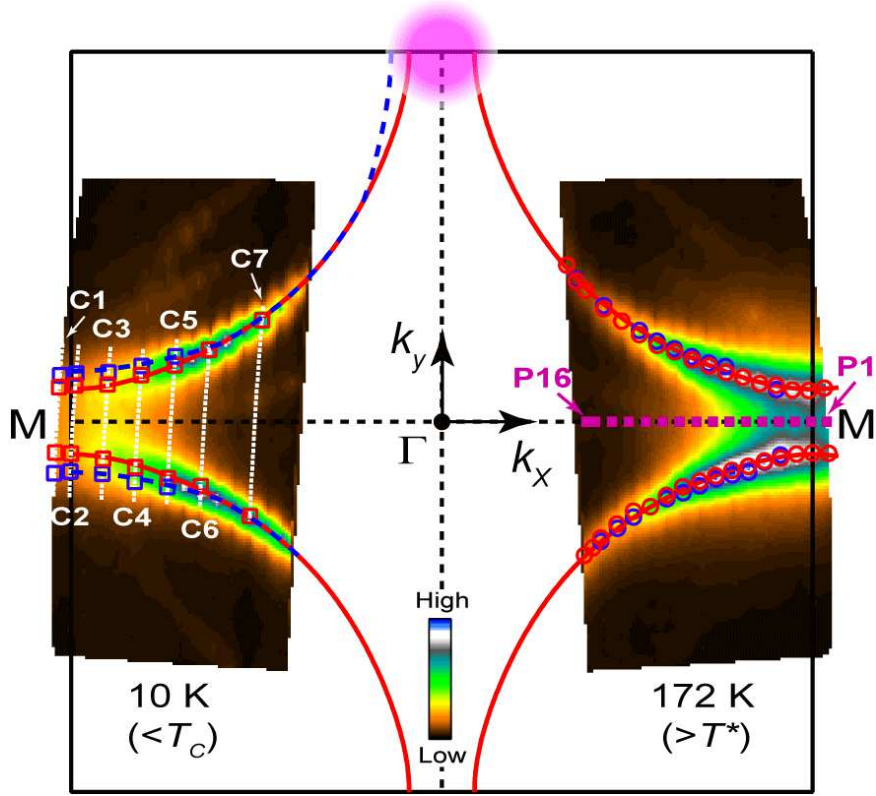}
\caption[ARPES Fermi surface maps] {Fermi surface maps measured below $T_c$ at 10 K (left) and above $T^*$ at 172 K (right) in the same momentum-space region (flipped for display). Dashed white lines labeled C1-C7 depict the cuts along which the EDCs shown in Fig. \ref{Fig. 2}, A-N were measured. Purple dots labeled P1-P16 along M-$\Gamma$ indicate momenta where EDCs in Fig. \ref{Fig. 2}, V-W) were measured. Red and blue squares on the left indicate momenta of the Fermi crossing $k_F$ ($k_{F1}$ and $k_{F2}$ in Fig. \ref{Fig. 2}, A-G) at 172 K and back-bending $k_G$ (black arrows in Fig. \ref{Fig. 2}, O-S) at 10 K of the dispersion of the EDC maximum along cuts C1-C7. Red and blue circles on the right indicate momenta of identifiable peaks in the momentum distribution curves (measured along cuts parallel to Cut C7) at $E_F$ at 172 K and 10 K, respectively. The solid red curves are a guide to the eye for the red squares and circles, whereas the dashed blue curve is the guide for the blue squares, together they show an increased $k_G-k_F$ misalignment going away from the nodal towards the antinodal region. The magenta-shaded region is approximately where multiple EDC features are found at 10 K (Figs. \ref{Fig. 2}W and S2F).}
\label{Fig. 1}
\end{figure}

\begin{figure}
\hspace*{-0.5cm}
\includegraphics [width=6.8in]{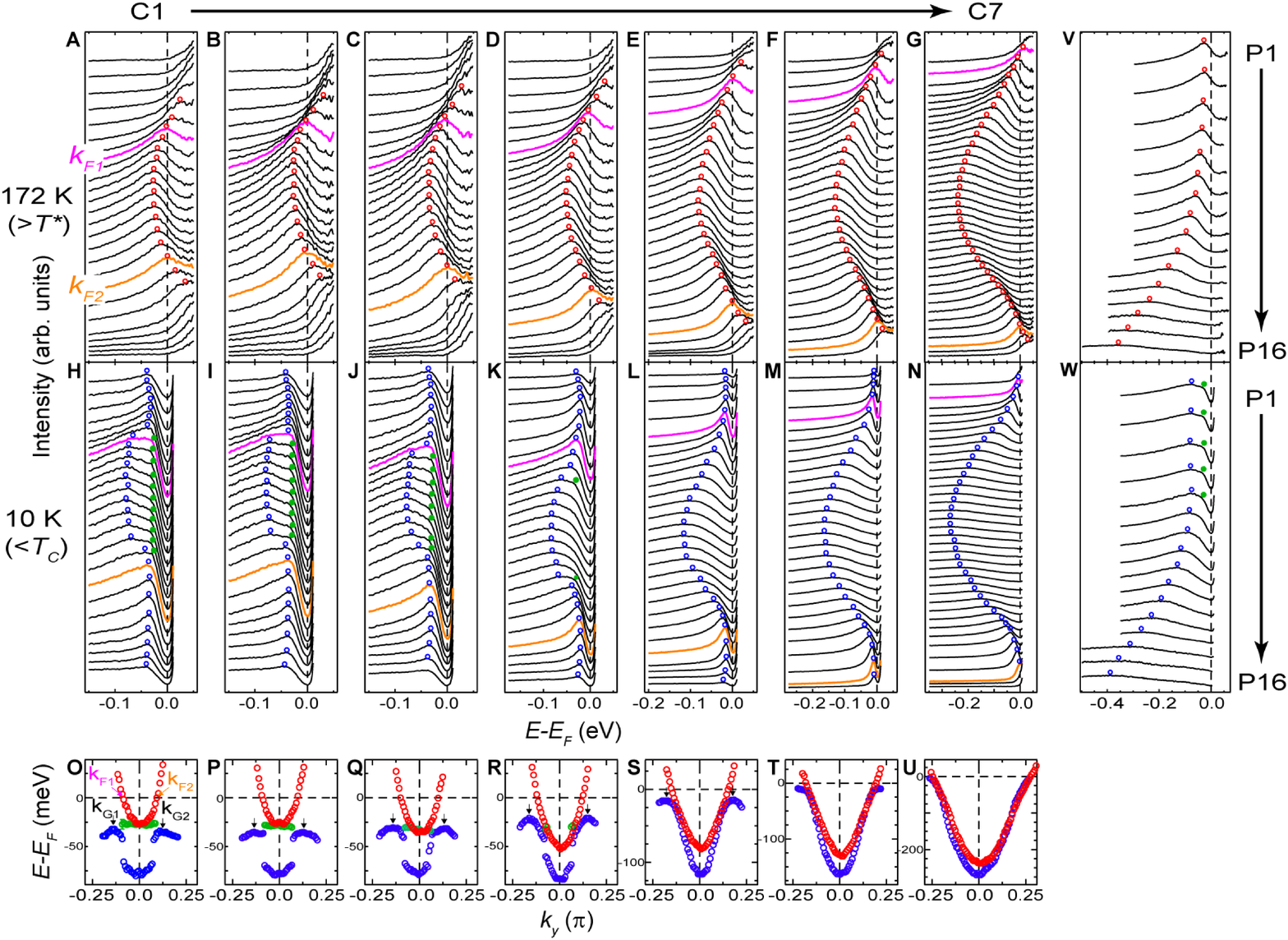}
\caption[ARPES momentum dependence above $T^*$ and below $T_c$] {(\textbf{A})-(\textbf{G}) and (\textbf{H})-(\textbf{N}) Selected EDCs at 172 K and at 10 K, respectively, for cuts C1-C7 approximately perpendicular to $\Gamma$-M (Fig. \ref{Fig. 1}). Each EDC corresponds to a white point in the cuts in Fig. \ref{Fig. 1}. EDCs in magenta and orange are located close to $k_F$. (\textbf{O})-(\textbf{U}) Dispersions of the EDC features in (A)-(N) for cuts C1-C7. For each dispersion curve, every other symbol corresponds to an EDC in (A)-(N). Error bars are estimated based on the sharpness of features, to be $\pm 3$ meV minimum and $\pm 8$ meV maximum [examples shown in (O)] based on various EDC analyses \cite{Note_SOM}. (\textbf{V})-(\textbf{W}) EDCs at momenta P1-P16 along M-$\Gamma$ (Fig. \ref{Fig. 1}) at 172 K and 10 K, respectively. Circles denote the EDC shoulder feature (solid green), the EDC maximum feature at 10 K (blue) and at 172 K (red).}
\label{Fig. 2}
\end{figure}

We will first analyze our ARPES data collected in different temperature regions. Above $T^*$, Pb-Bi2201 has a simple one-band band structure (right side of Fig. \ref{Fig. 1}). For each cut in momentum space perpendicular to $\Gamma$-M [$(0,0)$-$(\pi,0)$] (C1-C7 in Fig. \ref{Fig. 1}), the only distinct feature in the corresponding Fermi-function-divided \cite{Note_SOM} energy distribution curves (EDCs)  is a maximum (red circles in Fig. \ref{Fig. 2}, A-G). As a function of the $y$ component of the wave vector ($k_y$), the maxima have an approximately parabolic dispersion for each cut (red circles in Fig. \ref{Fig. 2}, O-U); the band bottom lies on the $\Gamma$-M axis, and the dispersion crosses the Fermi level ($E_F$) at two momenta, $k_{F}$ ($k_{F1}$ and $k_{F2}$). The band-bottom binding energy monotonically decreases from near $\Gamma$ to M (Fig. \ref{Fig. 2}, O-U). We take the Fermi-level crossings of this single band to define the Fermi surface. Despite the simplicity of the electronic structure above $T^*$, the width and energy-dependent broadening of the EDC maximum features, along with the familiar strange metal behavior seen in transport, imply that the system is not well described as a Fermi liquid.

We now turn to the temperature region below $T_c$. Here, the entire Fermi surface is gapped except at the nodal points ($k_{F}$ lying on the zone diagonal). In the nodal region, consistent with previous reports \cite{cuprates:Bi2201:TakahashiPGSC, cuprates:Bi2201:OneGap_Xingjiang, HTSC:Kondo_TwoPGs, cuprates:Bi2201:DingHongTwoGaps}, a $d$-wave-like gap along the Fermi surface is observed which we quantify as the energy position of the EDC maximum (blue circles) at $k_F$ (Fig. \ref{Fig. 2}, L-N). This maximum is still the only identifiable feature in the EDC. By comparing the EDCs in Fig. \ref{Fig. 2}, E-G with those in Fig. \ref{Fig. 2}, L-N, we see that the peaks of EDCs near $k_F$ are much sharper below $T_c$ than above $T^*$, however (perhaps surprisingly) the peaks well away from $k_F$ appear broader but with larger experimental uncertainties (also see Fig. \ref{Fig. 2}, V-W).

Away from the nodal region, the dispersion along each cut rises to a minimum binding energy and then bends back (Fig. \ref{Fig. 2}, H-K). These back-bendings (black arrows in Fig. \ref{Fig. 2}, O-S) occur at momenta $k_{G}$ ($k_{G1}$ and $k_{G2}$), which are increasingly separated from the Fermi surface (compare blue and red squares on the left side of Fig. \ref{Fig. 1}) towards the antinodes ($k_{F}$ lying on the zone boundary). Note that, for a superconducting gap, as a consequence of the particle-hole symmetry, one would expect $k_G\cong k_F$ (Fig. S6), as is the case in the nodal region. The substantial $k_G-k_F$ misalignment, previously found in a single antinodal cut perpendicular to $\Gamma$-M, has been interpreted as a signature of non-particle-hole symmetric order \cite{HTSC:PseudogapDispersion}. Our observation here complements that finding by providing an entire momentum-space picture.

The width and shape of the EDC near $k_F$ change fairly abruptly as a function of position along the Fermi surface, in contrast with the smooth evolution seen above $T^*$ (compare magenta curves in Fig. \ref{Fig. 2}, H-N and A-G). Strikingly, the antinodal EDC maxima below $T_c$ are broader than those above $T^*$ (Figs. \ref{Fig. 2}, V-W and S7I). Such broadening of the antinodal spectra with decreasing temperature, as reported previously \cite{HTSC:PseudogapDispersion}, is likely intrinsic, as the expected sharpening is observed simultaneously in the nodal spectra at low energy.

Around the M point, both the EDC line shape and the dispersion of the EDC maxima are more complex than their counterparts above $T^*$. The dispersion of the EDC maxima has two separate branches, one at relatively low energy which shows back-bendings at $k_G$ and the other at higher energy around the band bottom on the $\Gamma$-M axis. As one moves from cuts C1 to C7, the apparent discontinuity between the two branches near the antinode vanishes and the two branches merge (Fig. \ref{Fig. 2}, O-U). Both $k_G$ (blue squares in Fig. \ref{Fig. 1}) and the band-bottom energy (blue circles in Fig. \ref{Fig. 2}W) evolve smoothly, even as the dispersion at intermediate energies changes dramatically.  

Along with the changes in the dispersion of the EDC maxima below $T_c$, a well-defined shoulder feature (green dots in Fig. \ref{Fig. 2}, H-K and W) \cite{Note_SOM} emerges at low energy in the EDCs. This feature exhibits little dispersion either along $\Gamma$-M (Fig. \ref{Fig. 2}W) or perpendicular to it (Fig. \ref{Fig. 2}, H-J). Unlike the EDC maxima, its dispersion does not continue towards the zone center (Fig. \ref{Fig. 2}W) but instead loses its definition away from the vicinity of the M point (magenta-shaded region in Fig. \ref{Fig. 1}).
 
The changes in the spectral function on cooling do not occur smoothly; instead, abrupt changes in the thermal evolution occur in the neighborhoods of $T^*$ and $T_c$.

\begin{figure}
\centering
\includegraphics [width=5.5in]{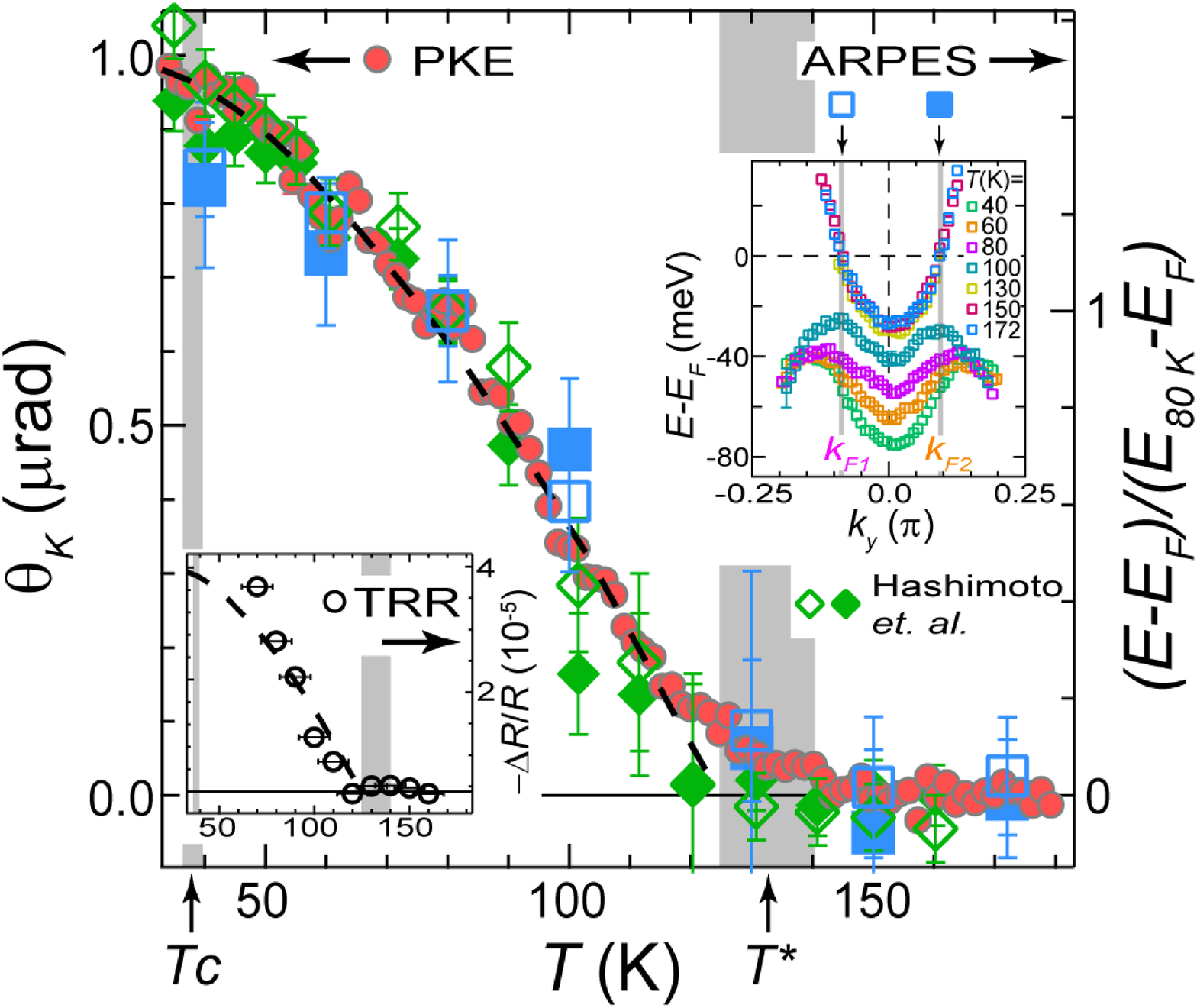}
\caption[Comparison of ARPES, PKE and TRR results] {Temperature dependence of Kerr rotation ($\theta_K$) measured by PKE, in comparison with that of the binding energy position of the EDC maximum at $k_F$ given by ARPES (reproduced from Fig. S1F and Ref. \cite{HTSC:PseudogapDispersion}). ARPES results are normalized to the 80 K values (free from the interference of fluctuating superconductivity). The dashed black curve is a guide to the eye for the PKE data, showing a mean-field-like critical behavior close to $T^*$ (see additional discussion \cite{Note_SOM}). Left inset: Temperature dependence of the transient reflectivity change measured by TRR (right axis). The dashed black curve (left axis) is reproduced from the main panel. Error bars (if invisible) are smaller than the symbol size. Right inset: The dispersion of the EDC maximum at various temperatures above $T_c$, summarizing the results of Figs. \ref{Fig. 2}A, S1, A-E and \ref{Fig. 4}A. All data were taken on samples from the same growth and annealing batch, except those reproduced from Ref. \cite{HTSC:PseudogapDispersion} on differently annealed samples.}
\label{Fig. 3}
\end{figure}

A detailed temperature-dependent study of an antinodal cut (C1 in Fig. \ref{Fig. 1}) shows that the dispersion of the EDC maxima (blue circles in Fig. S1, A-E) exhibit a transformation which begins at a temperature equal (within experimental uncertainty) to the reported values of $T^*$ in the literature \cite{Note_SOM}. Specifically, the energies of the EDC maxima at the band bottom and at $k_F$ are temperature independent above $T^*$ but begin shifting to higher binding energies below $T^*$ (Fig. S1F). The energy position at $k_F$ is a measure of the pseudogap, and can be defined as a spectral order parameter that becomes non-zero below $T^*$. Moreover, at temperatures below $T^*$ but well above $T_c$, one can already see the $k_G-k_F$ misalignment developing (right inset of Fig. \ref{Fig. 3}). These observations are consistent with our previous report on a similar cut in an OP Pb-Bi2201 sample prepared under a different post-annealing condition \cite{HTSC:PseudogapDispersion}.

It would be natural (as discussed below) to associate the abrupt change in nature of the ARPES spectra with a transition to a broken-symmetry state below $T^*$. To test this idea, we performed PKE measurements on the same crystals with finely-spaced temperature steps \cite{Note_SOM}. The results (Fig. \ref{Fig. 3}) are clearly suggestive of a slightly-rounded phase transition at $T^*$ below which a finite Kerr rotation emerges. A similar transition in the Kerr rotation was previously observed in high-quality YBa$_2$Cu$_3$O$_{6+x}$ (YBCO) crystals and was suggested to be linked to a broken-symmetry state that is not necessarily magnetic in nature \cite{HTSC:KerrEffect_YBCO}. The smallness of the previously observed Kerr rotation in YBCO suggested that it might not reflect the primary order. The present data again show a small Kerr rotation; however, the strong correspondence between the ARPES and PKE data in Fig. \ref{Fig. 3}, allows us to conjecture a phase transition at $T^*$, and also to corroborate the interpretation of previously published data on YBCO in terms of a similar transition.

We also studied TRR on the same Pb-Bi2201 crystals \cite{Note_SOM}. At temperatures between $T_c$ and $T^*$, we observed a negative signal with a time-resolution-limited turn-on followed by decay on the order of 100 femtoseconds (Fig. S3). Additionally, we observed a positive signal that emerges below $T_c$ and decays on a picosecond timescale. These observations are consistent with previous studies of various cuprate families that revealed different dynamics above and below $T_c$ \cite{HTSC:Pumpprobe_SCPG_1}, and associated the former with the pseudogap state \cite{HTSC:Pumpprobe_SCPG_2, HTSC:Pumpprobe_SCPG_3, HTSC:Pumpprobe_SCPG_YBCO, HTSC:Pumpprobe_pairing, HTSC:Pumpprobe_SCPG_Tl2223, HTSC:Pumpprobe_PG_2}. The striking aspects of the TRR results reported here are the clarity of the onset at $T^*$ and the direct correspondence with ARPES and PKE. As shown in the left inset of Fig. \ref{Fig. 3}, the magnitude of the negative TRR signal tracks the ARPES and PKE data quite well. This added correspondence further supports the existence of a phase transition at $T^*$ to the pseudogap state with near-equilibrium dynamics distinct from superconductivity.

\begin{figure}
\hspace*{-0.5cm}
\includegraphics [width=6.8in]{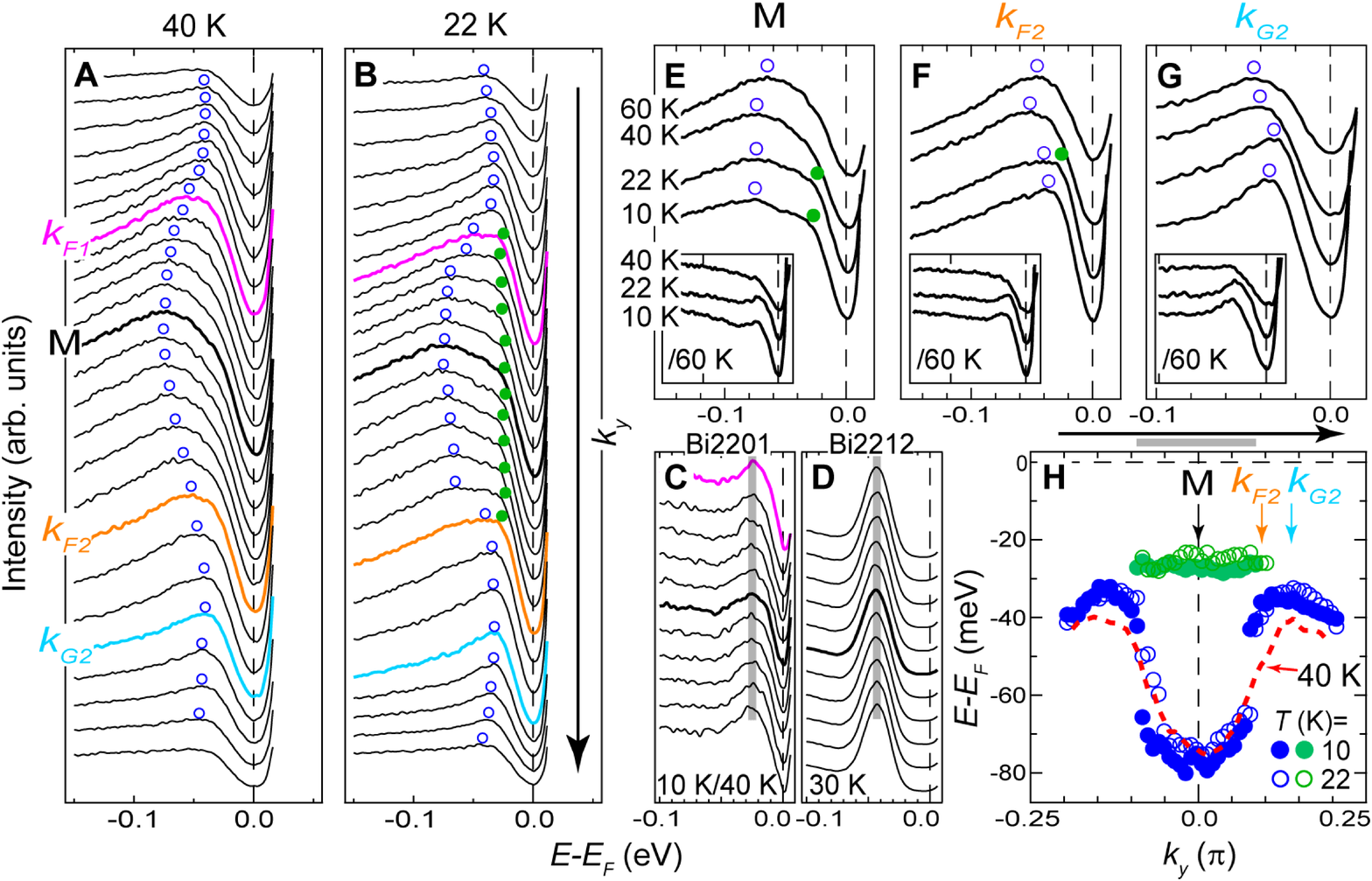}
\caption[ARPES temperature dependence across $T_c$] {(\textbf{A}) and (\textbf{B}) Selected EDCs at 40 K and 22 K along Cut C1 (Fig. \ref{Fig. 1}). See Fig. \ref{Fig. 2}, A and H for data at 172 K and 10 K, Fig. S1, A-E for other intermediate temperatures. (\textbf{C}) Antinodal EDCs at 10 K after dividing by the 40 K counterparts, covering the momentum range indicated by the grey bar in (H), in comparison with those in (\textbf{D}) taken in a similar range at 30 K on an OP Bi2212 sample. Non-dispersive peaks are seen in both cases despite different sharpness and energy positions. (\textbf{E})-(\textbf{G}) EDCs at different fixed momenta [specified in (A) and (H)] and temperatures around $T_c$. The counterintuitive increase of the antinodal gap, defined by the energy position of the EDC maximum in (F) and (G), with temperature rising above $T_c$ cannot be understood with a single energy scale assumed. Insets: Corresponding EDCs divided by the 60 K counterpart, showing the peaks losing definition above $T_c$ (Fig. S2E). (\textbf{H}) Summary for the dispersions of related EDC features across and below $T_c$. Vertical arrows specify momenta M, $k_{F2}$ at 172 K and $k_{G2}$ at 10 K. Apparent asymmetry of the dispersions across M is due to a finite deviation of the cut from the high-symmetry direction and a subtle balance of spectral weight between different features in the EDC. All EDC features and error bars are similarly determined as in Fig. \ref{Fig. 2}.}
\label{Fig. 4}
\end{figure}

Upon cooling below $T_c$, it is observed in ARPES that the shoulder on the low-energy side of the EDC maximum appears to develop somewhat above $T_c$ but well below $T^*$, and it grows truly distinct only below $T_c$ (compare Figs. \ref{Fig. 4}, A-B and S1, A-E). In order to obtain a clearer view of the structure of this feature, we show, in the insets of Fig. \ref{Fig. 4}, E-G, the spectra at low temperatures divided by the one at 60K - a procedure that converts the shoulder into a small peak. In Fig. \ref{Fig. 4}C we adopt the same procedure, but dividing by the 40K spectra instead, and in Fig. S2 we have subtracted off an approximate background from the EDCs. All three methods of analysis produce qualitatively similar results, with a small peak at a position that does not disperse appreciably along the exemplary antinodal Cut C1. In all cases, the energy and width of the peak do not change significantly with increasing temperature, while the peak intensity is strongly temperature-dependent; the peak becomes undetectable at a temperature which, while greater than $T_c$, is nowhere near $T^*$. This behavior of the processed data is highly reminiscent of the behavior of the superconducting coherence peak in Bi2212 \cite{HTSC:Pseudogap_ARPES_Fedorov}, although that peak is visible in the raw data (Fig. \ref{Fig. 4}D).

The strong and analogous temperature dependences of the ARPES, PKE and TRR data seen between $T_c$ and $T^*$ are most naturally understood if $T^*$ is associated with a phase transition to a non-superconducting broken-symmetry state \cite{Note_SOM}. While superconducting fluctuations are observable above $T_c$ (Figs. \ref{Fig. 4}, E-G and S2E), they appear to peter out at too low temperatures (being negligible already at 60 K) to play a central role in this higher temperature transition. To complete this picture, it is necessary to identify the nature of the broken-symmetry state, and to relate it to the apparently similar electronic changes that occur below $T^*$ in other cuprates \cite{Note_SOM}.

Below $T_c$, the nodal arc is gapped with a $d$-wave-like structure suggestive of a dominantly superconducting origin \cite{Note_nodal}. In contrast, in the antinodal region, rather than one order being dominant, or the two gaps of both orders adding in quadrature, the spectral function develops a complex structure with two energy scales below $E_F$ of mixed origin, a larger one being primarily associated with the pseudogap order, and a smaller one with the superconducting order.

To see what can be learned about the pseudogap order from ARPES, we have used a simple mean-field model \cite{Note_SOM} to compute the expected changes to the band structure induced by various forms of density wave \cite{stripe:theory:KivelsonStripe_eLiquidXtal, HTSC:CompetingOrders_DDW, stripe:theory:GrilliFluctuatingDW, stripe:neutron:TranquadaStripe1st, stripe:STM:CB_VortexCore, stripe:STM:CB_CB2212_Kapitulnik, stripe:STM:CB_Pseudogap, stripe:STM:C4C2, stripe:STM:PseudogapFluctStripe_Bi2212, stripe:STM:CB_Bi2201} (Fig. S7) or nematic order \cite{stripe:theory:KivelsonStripe_eLiquidXtal, stripe:STM:CB_CB2212_Kapitulnik, stripe:STM:C4C2, stripe:neutron:YBCO_eLiquidCrystal, HTSC:PseudogapBreakRotational, stripe:STM:Nematic_Bi2212} (Fig. S8) coexisting with $d$-wave superconductivity. We note that some (but certainly not all) key aspects of our experimental observations can be qualitatively reproduced, regardless of which of these orders is assumed \cite{Note_SOM}. An essential feature of both the experiment and the fits is the comparable sizes of the superconducting gap and the pseudogap. This implies that the two orders may have a more intimate connection than just competing orders, such as seen in $2H$-NbSe$_2$ where the charge density wave gap is at least 3 times the superconducting gap \cite{CDW:NdSe2_ARPES_twogaps}.



\clearpage

\begin{thebibliography}{1}

\bibitem{misc:PES:ZXreview}
A. Damascelli, Z. Hussain, Z.-X. Shen, Angle-resolved photoemission studies of the cuprate superconductors. \emph{Rev. Mod. Phys.} \textbf{75}, 473 (2003).

\bibitem{HTSC:PrecursorPairing}
V. J. Emery, S. A. Kivelson, Importance of phase fluctuations in superconductors with small superfluid density. \emph{Nature} \textbf{374,} 434 (1995).

\bibitem{HTSC:NernstEffect_Ong}
Y. Wang, L. Li, N. P. Ong, Nernst effect in high-$T_c$ superconductors. \emph{Phys. Rev. B} \textbf{73,} 024510 (2006).

\bibitem{cuprates:Bi2201:TakahashiPGSC}
K. Nakayama {\sl et al.}, Evolution of a pairing-induced pseudogap from the superconducting gap of (Bi, Pb)$_2$Sr$_2$CuO$_6$. \emph{Phys. Rev. Lett.} \textbf{102,} 227006 (2009).

\bibitem{cuprates:Bi2201:OneGap_Xingjiang}
J. Meng {\sl et al.}, Monotonic $d$-wave superconducting gap of the optimally doped Bi$_2$Sr$_{1.6}$La$_{0.4}$CuO$_6$ superconductor by laser-based angle-resolved photoemission spectroscopy. \emph{Phys. Rev. B} \textbf{79,} 024514 (2009).

\bibitem{stripe:theory:KivelsonStripe_eLiquidXtal}
S. A. Kivelson, E. Fradkin, V. J. Emery, Electronic liquid-crystal phases of a doped Mott insulator. \emph{Nature} \textbf{393,} 550 (1998).

\bibitem{HTSC:CompetingOrders_DDW}
S. Chakravarty, R. B. Laughlin, D. K. Morr, C. Nayak, Hidden order in the cuprates. \emph{Phys. Rev. B} \textbf{63,} 094503 (2001).

\bibitem{stripe:theory:GrilliFluctuatingDW}
M. Grilli, G. Seibold, A. Di Ciolo, J. Lorenzana, Fermi surface dichotomy in systems with fluctuating order. \emph{Phys. Rev. B} \textbf{79,} 125111 (2009).

\bibitem{HTSC:CompetingOrders_VarmaLoops}
C. M. Varma, Non-Fermi-liquid states and pairing instability of a general model of copper oxide metals. \emph{Phys. Rev. B} \textbf{55,} 14554 (1997).

\bibitem{cuprates:Bi2212:Kiyo_TwoGap}
K. Tanaka {\sl et al.}, Distinct Fermi-momentum-dependent energy gaps in deeply underdoped Bi2212. \emph{Science} \textbf{314,} 1910 (2006).

\bibitem{HTSC:Kondo_TwoPGs}
T. Kondo {\sl et al.}, Disentangling Cooper-pair formation above $T_c$ from the pseudogap state in the cuprates. \emph{Nat. Phys.} \textbf{7,} 21 (2011).

\bibitem{cuprates:Bi2201:DingHongTwoGaps}
J.-H. Ma {\sl et al.}, Coexistence of competing orders with two energy gaps in real and momentum space in the high temperature superconductor Bi$_2$Sr$_{2-x}$La$_x$CuO$_{6+\delta}$. \emph{Phys. Rev. Lett.} \textbf{101,} 207002 (2008).

\bibitem{stripe:neutron:TranquadaStripe1st}
J. M. Tranquada, B. J. Sternlieb, J. D. Axe, Y. Nakamura, S. Uchida, Evidence for stripe correlations of spins and holes in copper oxide superconductors. \emph{Nature} \textbf{375,} 561 (1995).

\bibitem{stripe:STM:CB_VortexCore}
J. E. Hoffman {\sl et al.}, A four unit cell periodic pattern of quasi-particle states surrounding vortex cores in Bi$_2$Sr$_2$CaCu$_2$O$_{8+\delta}$. \emph{Science} \textbf{295,} 466 (2002).

\bibitem{stripe:STM:CB_CB2212_Kapitulnik}
C. Howald, H. Eisaki, N. Kaneko, A. Kapitulnik, Coexistence of periodic modulation of quasiparticlestates and superconductivity in Bi$_2$Sr$_2$CaCu$_2$O$_{8+\delta}$. \emph{Proc. Natl. Acad. Sci.} \textbf{100,} 9705 (2003).

\bibitem{stripe:STM:CB_Pseudogap}
M. Vershinin {\sl et al.}, Local ordering in the pseudogap state of the high-$T_c$ superconductor Bi$_2$Sr$_2$CaCu$_2$O$_{8+\delta}$. \emph{Science} \textbf{303,} 1995 (2004).

\bibitem{stripe:STM:C4C2}
Y. Kohsaka {\sl et al.}, An intrinsic bond-centered electronic glass with unidirectional domains in underdoped cuprates. \emph{Science} \textbf{315,} 1380 (2007).

\bibitem{stripe:STM:PseudogapFluctStripe_Bi2212}
C. V. Parker {\sl et al.}, Fluctuating stripes at the onset of the pseudogap in the high-Tc superconductor Bi$_2$Sr$_2$CaCu$_2$O$_{8+x}$. \emph{Nature} \textbf{468,} 677 (2010).

\bibitem{stripe:STM:CB_Bi2201}
W. D. Wise {\sl et al.}, Charge density wave origin of cuprate checkerboard visualized by scanning tunneling microscopy. \emph{Nature Phys.} \textbf{4,} 696 (2008).

\bibitem{stripe:neutron:YBCO_eLiquidCrystal}
V. Hinkov {\sl et al.}, Electronic liquid crystal state in the high-temperature superconductor YBa$_2$Cu$_3$O$_{6.45}$. \emph{Science} \textbf{319,} 597 (2008).

\bibitem{HTSC:PseudogapBreakRotational}
R. Daou {\sl et al.}, Broken rotational symmetry in the pseudogap phase of a high-$T_c$ superconductor. \emph{Nature} \textbf{463,} 519 (2010).

\bibitem{stripe:STM:Nematic_Bi2212}
M. J. Lawler {\sl et al.}, Intra-unit-cell electronic nematicity of the high-$T_c$ copper-oxide pseudogap states. \emph{Nature} \textbf{466,} 347 (2010).

\bibitem{HTSC:BreakTimeRev_ARPES_Bi2212}
A. Kaminski {\sl et al.}, Spontaneous breaking of time-reversal symmetry in the pseudogap state of a high-$T_c$ superconductor. \emph{Nature} \textbf{416,} 610 (2002).

\bibitem{HTSC:BreakTimeRev_neutron_YBCO}
B. Fauqu$\acute{e}$ {\sl et al.}, Magnetic order in the pseudogap phase of high-$T_c$ superconductors. \emph{Phys. Rev. Lett.} \textbf{96,} 197001 (2006).

\bibitem{HTSC:KerrEffect_YBCO}
J. Xia {\sl et al.}, Polar Kerr-effect measurements of the high-temperature YBa$_2$Cu$_3$O$_{6+x}$ superconductor: Evidence for broken symmetry near the pseudogap temperature. \emph{Phys. Rev. Lett.} \textbf{100,} 127002 (2008).

\bibitem{HTSC:BreakTimeRev_neutron_Hg1201}
Y. Li {\sl et al.}, Unusual magnetic order in the pseudogap region of HgBa$_2$CuO$_{4+\delta}$. \emph{Nature} \textbf{455,} 372 (2008).

\bibitem{Note_SOM}
Materials and methods are available as supporting material on \textit{Science} Online.

\bibitem{otheroxides:KerrEffect_Sr2RuO4}
J. Xia, Y. Maeno, P. T. Beyersdorf, M. M. Fejer, A. Kapitulnik, High resolution polar Kerr effect measurements of Sr$_2$RuO$_4$: Evidence for broken time-reversal symmetry in the superconducting state. \emph{Phys. Rev. Lett.} \textbf{97,} 167002 (2006)

\bibitem{HTSC:PseudogapDispersion}
M. Hashimoto {\sl et al.}, Particle-hole symmetry breaking in the
pseudogap state of Bi2201. \emph{Nat. Phys.} \textbf{6,} 414 (2010).

\bibitem{HTSC:Pumpprobe_SCPG_1}
G. L. Eesley, J. Heremans, M. S. Meyer, G. L. Doll, S. H. Liou, Relaxation time of the order parameter in a high-temperature superconductor. \emph{Phys. Rev. Lett.} \textbf{65,} 3445 (1990).

\bibitem{HTSC:Pumpprobe_SCPG_2}
P. Gay {\sl et al.}, Femtosecond dynamics of BSCCO-2212. \emph{J. Low Temp. Phys.} \textbf{117,} 1025 (1999).

\bibitem{HTSC:Pumpprobe_SCPG_3}
J. Demsar, B. Podobnik, V. V. Kabanov, Th. Wolf, D. Mihailovic, Superconducting gap $\Delta_c$, the pseudogap $\Delta_p$, and pair fluctuations above $T_c$ in overdoped Y$_{1-x}$Ca$_x$Ba$_2$Cu$_3$O$_{7-\delta}$ from femtosecond time-domain spectroscopy. \emph{Phys. Rev. Lett.} \textbf{82,} 4918 (1999).


\bibitem{HTSC:Pumpprobe_SCPG_YBCO}
R. A. Kaindl {\sl et al.}, Ultrafast mid-infrared response of YBa$_2$Cu$_3$O$_{7-\delta}$. \emph{Science} \textbf{287,} 470 (2000). 

\bibitem{HTSC:Pumpprobe_pairing}
N. Gedik {\sl et al.}, Single-quasiparticle stability and quasiparticle-pair decay in YBa$_2$Cu$_3$O$_{6.5}$. \emph{Phys. Rev. B} \textbf{70,} 014504 (2004).

\bibitem{HTSC:Pumpprobe_SCPG_Tl2223}
E. E. M. Chia {\sl et al.}, Observation of competing order in a high-$T_c$ superconductor using femtosecond optical pulses. \emph{Phys. Rev. Lett.} \textbf{99,} 147008 (2007). 

\bibitem{HTSC:Pumpprobe_PG_2}
Y. H. Liu {\sl et al.}, Direct observation of the coexistence of the pseudogap and superconducting quasiparticles in Bi$_2$Sr$_2$CaCu$_2$O$_{8+y}$ by time-resolved optical spectroscopy. \emph{Phys. Rev. Lett.} \textbf{101,} 137003 (2008).

\bibitem{HTSC:Pseudogap_ARPES_Fedorov}
A. V. Fedorov {\sl et al.}, Temperature dependent photoemission studies of optimally doped Bi$_2$Sr$_2$CaCu$_2$O$_8$. \emph{Phys. Rev. Lett.} \textbf{82,} 2179 (1999).

\bibitem{Note_nodal}
That this is a bulk superconducting effect is corroborated by the magnetic field-dependent suppression of the Knight shift (a measure of the density of states at $E_F$) seen below $T_c$ of Bi2201 in nuclear magnetic resonance \cite{cuprates:Bi2201:NMR_knightshift_PG}. The Knight shift was found to drop sharply at $T^*$, exhibiting a similar temperature dependence as those shown in Fig. \ref{Fig. 3}.

\bibitem{cuprates:Bi2201:NMR_knightshift_PG}
S. Kawasaki, C. Lin, P. L. Kuhns, A. P. Reyes, G.-q. Zheng, Carrier-concentration dependence of the pseudogap ground state of superconducting Bi$_2$Sr$_{2-x}$La$_x$CuO$_{6+\delta}$ revealed by $^{63,65}$Cu-nuclear magnetic resonance in very high magnetic fields. \emph{Phys. Rev. Lett.} \textbf{105,} 137002 (2010).

\bibitem{CDW:NdSe2_ARPES_twogaps}
S. V. Borisenko {\sl et al.}, Two energy gaps and Fermi-surface ~``arcs" in NbSe$_2$. \emph{Phys. Rev. Lett.} \textbf{102,} 166402 (2009).

\end{thebibliography}



\begin{scilastnote}
\item We thank I. Vishik, W.-S. Lee, L. Taillefer, M. Greven for helpful discussions, Y. Li and G. Yu for experimental assistance on SQUID and J.-H. Chu on resistivity measurements. R.-H.H. thanks the SGF for financial support. This work at SIMES and ALS is supported by the Department of Energy, Office of Basic Energy Sciences under contracts DE-AC02-76SF00515 and DE-AC02-05CH11231.
\end{scilastnote}


\clearpage

\end{document}